# Adaptive Neighboring Selection Algorithm Based on Curvature Prediction in Manifold Learning

Lin Ma[1], Caifa Zhou[2], Xi Liu[3], Yubin Xu[4]

1: Communication Research Center, Harbin Institute of Technology, Harbin, China, malin@hit.edu.cn
2: Communication Research Center, Harbin Institute of Technology, Harbin, China, albertchow@126.com
3: Communication Research Center, Harbin Institute of Technology, Harbin, China, liu9054@126.com
4: Communication Research Center, Harbin Institute of Technology, Harbin, China, ybxu@hit.edu.cn

*Abstract*—Recently manifold learning algorithm for dimensionality reduction attracts more and more interests, and various linear and nonlinear, global and local algorithms are proposed. The key step of manifold learning algorithm is the neighboring region selection. However, so far for the references we know, few of which propose a generally accepted algorithm to well select the neighboring region. So in this paper, we propose an adaptive neighboring selection algorithm, which successfully applies the LLE and ISOMAP algorithms in the test. It is an algorithm that could find the optimal K nearest neighbors of the data points on the manifold. And the theoretical basis of the algorithm is the approximated curvature of the data point on the manifold. Based on Riemann Geometry, Jacob matrix is a proper mathematical concept to predict the approximated curvature. By verifying the proposed algorithm on embedding Swiss roll from R3 to R2 based on LLE and ISOMAP algorithm, the simulation results show that the proposed adaptive neighboring selection algorithm is feasible and able to find the optimal value of K, making the residual variance is relatively small and better visualization of the results. By quantitative analysis, the embedding quality which is measured by residual variance increases 45.45% after using the proposed algorithm in LLE.

*Keywords- Manifold learning; Curvature prediction; Adaptive neighboring selection; Residual variance*

I. INTRODUCTION

Mapping the data from input space into low dimensional space is inevitable in solving many computer science problems, especially in computer vision and pattern recognition [1], with the development information technology. And manifold learning is a powerful method based on the cognitive psychology of human beings [2], and it could tackle the problem of discovering intrinsically low-dimensional structure embedded in high-dimensional data sets [3].

There are many methods for dimensionality reduction based on the assumption of manifolds. Different approaches concentrate on preserving various characteristics of the manifolds. In [4], Roweis suggested an algorithm named Local Linear Embedding (LLE), which was a method to preserve local linear structure of high-dimensional data sets when embedding data to a low-dimensional space. Isometric Mapping (Isomap) tries to preserve global geometry of the manifold[2]. And another method, named Laplacian Eigenmap (LE), is derived from LLE[5]. The above three methods have the same feature that the Euclidean distance is implemented to show the structure features of the manifold. According to whether the manifold learning is based on the assumption linear structure of high-dimensional data sets, LLE and LE are called linear dimensionality reduction, while Isomap is called nonlinear dimensionality reduction. Except for the mentioned three methods, there are also numerous other dimensionality reduction methods designed for specific purposes and applications, which are less popular and would not be taken into consideration in this paper.

Therefore, as we know that many of manifold learning methods are based on LLE, LE or Isomap [6]. And we get the information that manifold learning methods have similar or the same basic framework [5-7]. And neighboring selection is the first step in the framework. Generally, there are two strategies applied most. One is K nearest neighbors (KNN), the other is $\varepsilon$-ball [2,4]. Due to the KNN strategy is easily applied in the data sets which are not convergent, we analyze this strategy in this paper, and aim at proposing an algorithm which could adaptively select the optimal neighboring, not only locally but also globally, based on the features of manifolds.

When it comes to adaptive neighboring selection method of manifold learning, there are several previous researching results from the references which we read. In [7], Kouropteva proposed an automatic method for selecting of optimal parameters of LLE. But the method is only applied in LLE and it is a method that only could find a relative optimal fixed neighboring for the global data sets. In [8], the author suggested an adaptive neighboring selection method by considering about the local curvature of the manifold and calculating the relative changing of curvature of the manifold. And the author recommended that compression and expansion methods according to the relative curvature changing of the manifold. Although the method works when applied in Local Tangent Space Alignment (LTSA) and other several methods, the complexity of the algorithm, both time and space, increased greatly comparing to LTSA which had not add the adaptive neighboring selection method.

In this paper, we propose an adaptive neighboring selection method which could find, both local and global, optimal neighboring of the manifold and at the same time the increasing of complexity of algorithm is relative low at the same time. In additional, the proposed algorithm is compatible with many other manifold learning methods on condition that they have the same or similar basic framework.

The rest of this paper is as follows. Section II will analyze the curvature prediction of manifold and this is the key to finding

This paper is supported by National Natural Science Foundation and Civil Aviation Administration of China. (Grant No. 61101122 and 61071105) and Fundamental Research Funds for the Central Universities (Grant No. HIT. NSRIF.2010090) the adaptive neighboring selection method. Section III will investigate the accomplishment algorithm and the system error. The experiment result of adaptive neighboring selection algorithm which applies in Isomap and LLE will be shown in section IV and analyze. Finally, the conclusions will be drawn in section V.

## II. CURVATURE PREDICTION OF THE MANIFOLD

### A. Manifold Learning

Manifold learning is based on perceptive psychology of human beings and considering it from academic perspective, it means that closing neighboring region must have ample overlapping to keep and strengthen the efficiency of information transmitting on local manifold [9]. To keep the topology of the data sets stable, the value of K must be bigger than $d$, the dimension of low-dimensional space [10]. Thus, the inferior value of K is $d+1$.

Considering about the relationship between $K$ and the curvature of the manifold, the value of $K$ should decrease if the curvature of local manifold increases, otherwise, the value of $K$ could increase. Based on the assumption that the high-dimensional space is a smooth manifold, finding out a method to predict the curvature of manifold is a significant work. From Riemannian Differential Geometry, the curvature of one point of multi-factor function could compute by Jacobi matrix of the function [11]. Therefore, to calculate the curvature of the manifold, working out the functional relationship of input data sets is another key work. For having no idea about the number of the dependent variables and the independent variables, however, it is difficult to work out the functional relationship of the manifold. To achieve the accomplishment, an approximated method is proposed to estimate Jacobi matrix of discrete data sets in [10].

### B. Theoretical basic of curvature prediction

Assuming that the data points sampled from a smooth manifold $M = f(\Omega)$, where: $f: \Omega \subset R^d \to R^m$ is a smooth mapping that defines in open connected set $\Omega$. Considering a chosen point $x_i$ and its local structure on the manifold, having:

$$x_i = f(\tau) \quad (1)$$

Where $x_i$ is a point in data sets.

To the neighboring point of $x_i$, it could be expressed as follows:

$$\hat{x} = x + J_\tau \cdot (\hat{\tau} - \tau) + \varepsilon(\hat{\tau}, \tau) \quad (2)$$

where $J_\tau$ is the Jacobi matrix of point $\tau$.

From Riemann geometry, the derivation of multifactor function could be expressed by the determinant of Jacobi matrix as following equation:

$$\frac{\partial(y_1, y_2, \cdots, y_m)}{\partial(x_1, x_2, \cdots, x_m)} = \det\left(\frac{\partial y_i}{\partial x_j}\right) \quad (3)$$

Where $(\partial y_i / \partial x_j)$ is the Jacobi matrix of the function. Thus, the curvature of point $x_i$ could be calculated by Jacobi matrix. And the theoretical analyzing could see [8] and [11].

For having no accurate functional relationship of the manifold, neither have Jacobi matrix and data point $\tau$. By using the approach proposed in [11], Jacobi matrix could be estimated by the limited neighboring points of $x_i$. In this paper we assume that $N_i = \{x_{i1}, x_{i2}, \cdots, x_{iN}\}$ is the nearest neighboring set of $x_i = f(\tau_i)$. Using principal component analysis (PCA) and singular value decomposition (SVD), having the equation:

$$x_i - J_\tau \cdot (\hat{\tau} - \tau) = \overline{x}_i + Q_i \theta_j^{(i)} \quad (4)$$

Where $\overline{x}_i$ is the center of $N_i$; $Q_i$ is a matrix that consists by $r$ largest Eigen-vectors of SVD; $\theta_j^{(i)}$ is the Eigen-vector of SVD; SVD operated on matrix $[x_{i1} - \overline{x}_i, x_{i2} - \overline{x}_i, \cdots, x_{iN} - \overline{x}_i]$. And the theoretical analyzing details are in [11] and [12].

### C. Computation of Curvature

By the above analyzing, a significant problem could be tackled by Riemann Geometry and PCA and a probable approach for computing the curvature of point in data sets is proposed in following analysis. From Eq. (4), deducing equation:

$$\|J_\tau \cdot (\hat{\tau} - \tau)\| = \|\overline{x}_i - x_i + Q_i \theta_j^{(i)}\| \quad (5)$$

From Eq. (5), another equation is:

$$\|J_\tau\| \cdot \|(\hat{\tau} - \tau)\| = \|\overline{x}_i - x_i + Q_i \theta_j^{(i)}\| \quad (6)$$

Thus, following equation could be worked out:

$$\|J_\tau\| = \frac{\|\overline{x}_i - x_i + Q_i \theta_j^{(i)}\|}{\|\theta_j^{(i)}\|} \quad (7)$$

By simple deducing,

$$J_{\inf} = \frac{\|\overline{x}_i - x_i\| + \|Q_i \theta_j^{(i)}\|}{\|\theta_j^{(i)}\|} \quad (8)$$

In Eq. (8), $J_{\inf}$ is the inferior value of Jacobi matrix. Thus, the curvature of data points in the manifold could be estimated by (8). And there is another key work need to analysis is how to choose the number of neighboring region of $x_i$. A simple strategy is indicated in [13].

$$N = \begin{cases} 8, & \text{if D<d} \\ 4d, & \text{if D} \geq \text{d} \end{cases} \quad (9)$$

Therefore, computing the inferior value of Jacob matrix and treating it as the approximated curvature of points on the manifold.

## III. ADAPTIVE SELECTION OF K NEAREST NEIGHBORS

After working out the curvature of the point, but a significant problem is the relationship between the value of $K$ and the value of the curvature. First, the inferior value and the superior value of K is the limited range of $K$. From [10], the inferior value of

$K$ is $d+1$. And the superior value of $K$ is estimated by $6D$ [13]. And we compute adaptive selection of K by:

$$K_i = K_o + \text{int}\left[\frac{(\Delta J_\tau)}{\delta_o}\right] \quad (10)$$

Where: $K_o$ and $K_i$ is the initial value of K and K nearest neighbors of $x_i$, respectively. And $\Delta J_\tau$ is the ranging curvature of data points in one neighboring region. Another parameter is $\delta_o$ and it is a criterion to measure the changing of curvature to the changing of number of nearest neighbors. Thus, we have the adaptive neighboring selection rule:

$$K = \begin{cases} K_i, & \text{if } K_i \in [K_{\inf}, K_{\sup}] \\ K_{\inf}, & K_i < K_{\inf} \\ K_{\sup}, & K_i > K_{\sup} \end{cases} \quad (11)$$

Where, $K_i$ is defined as Eq. (10). By the rule, adaptive neighboring of every data point could be estimated. However, working out the initial value of K is difficult. From the consequence of several manifold learning algorithms, setting $K_o$ equals to 8, roughly.

After applying the adaptive neighboring selection algorithm in different manifold learning methods, an efficient method to measure which neighboring selection method is better or how measure the optimal result of the adaptive neighboring selection algorithm. In general, there are different definitions of "optimality". We rely on quantitative measures introduced below to characterize this term in order to avoid a subjective evaluation often accompanying a human visual check used in many cases.

We need to find out a method to estimate the "quality" of input-output mapping and work out a parameter which could represent how well the high-dimensional structure is mapped to the embedded space. In [4], the residual variance is regarded as a suitable parameter to fulfill this purpose. The computation of the residual variance, however, has different ways based on different ideas. And the most two common methods are introduced in the paper.

One method is based on the standard linear correlation coefficient. The value of residual variance equals to $1 - \rho^2_{D_X, D_Y}$, where $\rho$ is the standard linear correlation coefficient, taken over all entries of $D_X$ and $D_Y$; $D_X$ and $D_Y$ are the matrices of Euclidean distances (between pairs of points) in X and Y, respectively. X and Y are the input high-dimensional data set and output embedded low-dimensional data set, respectively. And we express the residual variance in following expression:

$$\xi_{r\text{var}} = 1 - \rho^2_{D_X, D_Y} \quad (12)$$

Another definition of the residual variance is based on the sum of residual Eigen values. In this, we use Eq. (10) to measure the quality of input-output mapping in Part 4 of the paper. Generally, the lower the residual variance is, the better high-dimensional data are represented in the embedded space.

For most typical manifold learning algorithms used stable value of K as nearest neighbors, we need to design the algorithm to improve the compatibility to present typical manifold learning algorithms.

According to the analyzing of Part 2 of this paper, we can describe the adaptive neighboring method as follows:

- Select $N$ according to Eq. (9)--the neighboring region of $x_i$, to compute the approximated curvature;
- Calculate $K_i$ according to Eq. (11)--the adaptive neighboring selection, and store in matrix $KAN$;
- Compute embedded consequence from X to Y by using LLE and Isomap algorithm, respectively according to $KAN$;
- Compute embedded consequence from X to Y by using LLE and Isomap algorithm, respectively using stable value of $K$;
- Calculate $\xi_{r\text{var}}$ of different algorithms and different parameters according to Eq. (12);

IV. IMPLEMENTTION AND PRFORMANCE ANALYSIS

We apply the adaptive neighboring selection algorithm in mapping Swiss roll into two-dimensional space by using LLE and Isomap algorithm. We analyze the existence of optimal K nearest neighbors of the algorithm. We embed 800 data points from $R^3$ to $R^2$ and calculate the residual variance of different value of K and figure out the changing tendency of residual variance with changing of K (Figure 1: residual variance changing tendency).

As Fig. 1 shows that there is an optimal value of K which could minimize the residual variance. We also find out that there is a fluctuation of residual variance with the changing of K. According to the basic principle of manifold learning, the value of K must be large enough to ensure the efficiency of information transmitting and when it comes to linear and local manifold learning algorithms, like LLE, etc, the value of K could not be too large, since as the increasing of K, the local linear structure of the manifold could not satisfy the requirements of corresponding algorithms. From the information of previous references, we have the idea that the value of K could not be too large for the complexity of the algorithm increases much faster as the increasing value of K. We find out the optimal value of K for LLE and Isomap algorithm, and the value of K equals to 14, 8, respectively from Fig. 1.

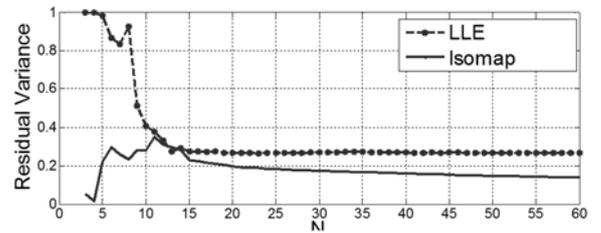

Figure 1. Residual Variance Tendency

Then we apply adaptive neighboring selection algorithm in the selecting of K in LLE and Isomap algorithm. In LLE algorithm, we could compute out the adaptive value of K is 10, which does not equal to the optimal value of LLE algorithm, but it is a inflection point of the curve. However, we can see that the residual variance of the algorithm might be small on the condition that the value of K is very small from Fig. 1. From visional result of manifold learning of Swiss roll, we can construe this paradox by visional figures of Swiss roll. We can see that the visual consequence is not good when the value of K is not large enough or too large (Figure 2: Visional Consequences of Manifold Learning of Swiss Roll). For we compare the visional results of Swiss roll with the original smooth manifold, the basic structure of it should be preserved when we apply manifold learning algorithm on it.

Figure 2 shows the results of testing LLE and Isomap on a Swiss roll data set. 800 points were generated uniformly in a rectangle (top left) and mapped into a Swiss roll configuration in $R^3$. LLE and Isomap recovers the rectangular structure correctly provided that the neighborhood parameter is not too large (in this case $K$ =8 and $K$ =10, respectively).

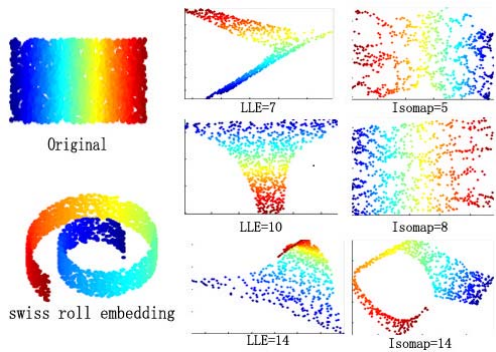

Figure 2. Manifold Learning of Swiss Roll

From quantitative analysis, we have the residual variance of LLE and Isomap which is shown in Table I with K=8, 10, 12, respectively.

TABLE I.  PART OF RESIDUAL VARIANCE

|        | *K=8*  | *K=10* | *K=12* |
|--------|--------|--------|--------|
| **LLE**    | 0.5132 | 0.4064 | 0.3318 |
| **Isomap** | 0.2788 | 0.2800 | 0.3098 |

After using the proposed algorithm in the paper, an optimal value of $K$ is 14 and 8 for LLE and Isomap, respectively. Thus, we have the optimal residual of variance of LLE and Isomap is 0.2799 and 0.2788, respectively. The relative increasing of embedding quality is 45.45% and 1.00% of LLE and Isomap, respectively. The relative increasing of embedding quality is calculated by $\left(\text{Resi}_{max} - \text{Resi}_{optimal}/\text{Resi}_{max}\right)$, where $\text{Resi}_{max}$ and $\text{Resi}_{optimal}$ stands for maximum value of residual variance in Table1 and optimal residual variance of LLE and Isomap, respectively.

We also compute out different value of K on the condition that dividing all data points into several groups. For the original data points are sampled from a smooth and continuous manifold, so we find out that the residual is smaller when we treat all data points in one group than the residual variance when we divide all data points into several parts. We can see the changing from figure 3, which indicates the residual variance changing tendency of group dividing. The parameter G stands for the number of groups divided during the test. And we find out that G equals to 1, the residual variance has minimum value. It might work when we apply adaptive neighboring selection algorithm to a normal manifold, non-smooth or non-continuous on the condition that we divide all data sets into several parts.

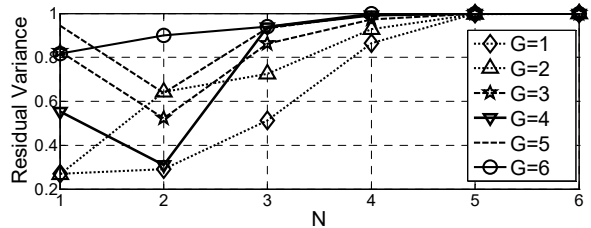

Figure 3.  Residual Variance Changing Tendency of Group Dividing

## V. CONCLUSION

The selection of neighboring region of manifold learning method is one of the key steps to achieve the high-dimensional smooth manifold embedded to low-dimensional space. We propose an adaptive neighboring selection algorithm which could select an optimal value of K depending on the curvature or the approximated curvature of data points on the manifold and we also consider about the embedded dimensions to determine the related parameter of adaptive neighboring selection algorithm. We test our algorithm by applying the adaptive neighboring selection algorithm in LLE and Isomap algorithm, and analyze the performance of it by compute and figure the curve of residual variance of embedding consequences. And we analyze visional results of manifold learning methods. We conclude that the adaptive neighboring selection algorithm works when applied to LLE and Isomap algorithm and it might be able to apply to other manifold learning methods and analyze the characteristic of non-smooth and non-continuous manifolds. And we would like to analyze several works in the future research.

## ACKNOWLEDGMENT

The author makes particular thanks to Zhongzhe Deng and Bin Luan, who help to realize the algorithm by MATLAB and edit part of picture in the paper. The author also wishes to thank all members in the program groups whose ideas are contributive to the paper.